\documentclass{aa}
\usepackage{graphicx}
\usepackage{txfonts}
\begin{document}
\title{SSSPM\,J1444$-$2019: an extremely high proper motion, 
       ultracool subdwarf
\thanks{Based on archival data from the SuperCOSMOS Sky Surveys, 2MASS, and 
DENIS, and observations with the ESO 3.6-m telescope (ESO 072.C-0630) and 
VLT (ESO 072.C-0725)}}

%   \title{SSSPM J1444$-$2019: an object of extremely high proper motion 
%         (3.5 arcsec/yr) classified as a nearby ultracool subdwarf 
%         \thanks{based on the systematic search in archival data 
%         from the SuperCOSMOS Sky Surveys, 2MASS and DENIS, 
%         and on spectroscopic observations with the ESO\,3.6-m telescope 
%         (ESO 072.C-0630) and the VLT (ESO 072.C-0725)}}

   \titlerunning{SSSPM\,J1444$-$2019: an extremely high proper motion, 
                 ultracool subdwarf} 

%   \subtitle{}

   \author{R.-D. Scholz
          \inst{1}
          \and
          N. Lodieu
          \inst{1,2}
          \and
          M. J. McCaughrean
          \inst{1,3}
          }

   \offprints{R.-D. Scholz}

   \institute{Astrophysikalisches Institut Potsdam, An der Sternwarte 16,
              14482 Potsdam, Germany \\
              \email{rdscholz@aip.de}
            \and 
              University of Leicester, Department of Physics \& Astronomy,
              University Road, Leicester LE1 7RH, UK \\
              \email{nl41@star.le.ac.uk}
            \and 
              University of Exeter, School of Physics, Stocker Road,
              Exeter EX4 4QL, UK \\
              \email{mjm@aip.de, mjm@astro.ex.ac.uk}
             }

   \date{Received ...; accepted ...}

\abstract{We present the discovery of a new extreme high proper motion object
($\sim$3.5 arcsec/year) which we classify as an ultracool subdwarf 
with [M/H]\,$\sim -0.5$. It has
a formal spectral type of sdM9 but also shows L-type features: while the VO 
bands are completely absent, it exhibits extremely strong TiO absorption in 
its optical spectrum. With a radial velocity of about $-160$\,km/s and a 
rough distance estimate of 16--24\,pc, it is likely one of the nearest halo 
members crossing the Solar neighbourhood with a heliocentric space velocity of
$(U,V,W)=(-244,-256,-100)\pm(32,77,6)$\,km/s.

   }

   \maketitle

%________________________________________________________________

\section{Introduction}
It is well understood that the observed magnitudes of stars cannot be taken
as direct indicator of their true luminosities without accounting for their
distances. Similarly, the distance must be known before their measured proper 
motions on the celestial sphere can be converted into their true velocities. 
Thus, high proper motion (hereafter HPM) objects represent a mixture of 
{\it very nearby} stars, neighbours of the Sun in the local Galactic spiral 
arm, and {\it very fast} representatives of the Galactic thick disk and halo, 
just passing through the neighbourhood. As a consequence, even though thick 
disk and halo stars are relatively rare when compared to the number density 
of stars in the Solar neighbourhood, they are over-represented in HPM samples
(Digby et al.\ \cite{digby03}). 

In HPM works such as the Luyten Half Second (LHS) catalogue 
(Luyten~\cite{luyten79}), the stellar proper motions are listed in units of 
arcsec/yr, even though the majority of stars in those catalogues have proper 
motions below 1~arcsec/yr. In the LHS, only about 530 sources exceed that 
limit, although an additional 36 have been discovered during the last 15 years,
suggesting that the sample is by no means yet complete (cf.\ Table~1 of
Subasavage et al.\ \cite{subasavage04}). At the more extreme end, there are
just 25 LHS stars with proper motions between 3.5 and 10.3~arcsec/yr, but
there again the sample is incomplete: after decades of stagnation, three 
new members have recently been added to this elite club, namely the red 
dwarf star SO~025300.5+165258 with $\mu$\,$\sim$\,5.1~arcsec/yr (Teegarden 
et al.\ \cite{teegarden03}) and the brown dwarf binary 
$\varepsilon$\,Indi\,Ba,Bb with $\mu$\,$\sim$\,4.7~arcsec/yr 
(Scholz et al.~\cite{scholz03};
McCaughrean et al.\ \cite{mjm04}). These latter two objects are also the 
first brown dwarfs known in the immediate Solar neighbourhood, i.e.\ within 
a horizon of about 4\,pc.

In this letter, we present the discovery of a new object with 
$\sim$3.5~arcsec/yr proper motion, fainter and redder than all other known 
extreme HPM objects apart from the brown dwarfs $\varepsilon$\,Indi\,Ba,Bb  
(Sect.~\ref{pmphot}). We describe low- and high-resolution spectroscopic 
follow-up observations which have allowed us to classify the object and to 
measure its radial velocity (Sect.~\ref{specobs}). Based on all the available 
data, we discuss the likely status of the object as one of the nearest cool 
halo subdwarfs (Sect.~\ref{concdisc}). 

%__________________________________________________________________

%%%%%%%%%%%%%%%%%%%%%%%%%%%%%%%%%%%%%%%%%%%%%%%%%%%%%%%%%%%%
%%%%%%%%%%%%%%%% PLOT OF 4 charts %%%%%%%%%%%%%%%%
%%%%%%%%%%%%%%%%%%%%%%%%%%%%%%%%%%%%%%%%%%%%%%%%%%%%%%%%%%%%
%
\begin{figure*}[htb]
\begin{center}
\caption[4 charts]{Finding charts of SSSPM J1444$-$2019 (circled) at different
epochs clearly illustrating its extremely large proper motion. The different 
passband data are sorted according to the epochs at which they were taken,
from blue and red SSS plates to the most recent epoch 2MASS $J$ band image,
and also demonstrate the very red optical and optical-to-infrared colours of 
the object (see also Table~\ref{sss2m}). North is up and East left. 
}
\label{4ima}
\end{center}
\end{figure*} 
 
\section{Proper motion measurement and photometry}
\label{pmphot}
SSSPM~J1444$-$2019 was discovered in the course of our ongoing HPM survey of 
the southern sky using multi-epoch optical data from the SuperCOSMOS Sky 
Surveys (SSS; Hambly et al.\ \cite{hambly01a,hambly01b,hambly01c}) and 
near-infrared data from the Two-Micron All Sky Survey (2MASS; Cutri et al.\
\cite{cutri03}). After the $\varepsilon$\,Indi\,Ba,Bb system it has the 
second largest proper motion among the objects we have found to date. The 
search strategy which led to the discovery of SSSPM J1444$-$2019 is briefly 
described here.

First, we selected bright ($J$\,$<$\,14) objects in the 2MASS database without 
an optical counterpart, working on the basis that an epoch difference of 
$\sim$15 years between the 2MASS observations and the optical A2.0 catalogue 
(Monet et al.\ \cite{monet98}) in the southern sky might mean the lack of 
cross-identification is caused by a large proper motion of the object. 
In order to exclude crowded fields and the Galactic plane, we restricted our
search to higher Galactic latitudes ($|b|$\,$>$\,20$^{\circ}$) and selected 
only isolated 2MASS sources with no neighbours inside a 30~arcsec radius.
About $\sim$\,11,000 candidates were extracted in this manner and this sample
was then checked for true HPM objects using SSS finding charts in three 
passbands at four epochs. The success rate was $\sim$3\,\%, with most of 
the confirmed objects turning out to be previously known HPM stars.
 
For each new HPM object, the accuracy of the proper motion determination was
improved by using all available SSS positions (including those from overlapping 
plates), the 2MASS position, and additional epoch data from the DEep 
Near-Infrared Survey (DENIS; Epchtein et al.\ \cite{epchtein97}) if available. 
The huge proper motion of SSSPM~J1444$-$2019 was reason alone for follow-up 
observations, but in addition, its very red optical and optical-to-infrared
colours but rather blue $J$--$K_s$ colour (Table~\ref{sss2m}) triggered
special attention and led us to speculate that it was a potential subdwarf. 

\begin{table*}
 \footnotesize
 \caption[]{Astrometry and photometry from SSS, 2MASS, and DENIS.
}
\label{sss2m}
 \begin{tabular}{lcccccccccc}
 \hline
\hline
Name & $\alpha, \delta$ & Epoch & $\mu_{\alpha}\cos{\delta}$ & $\mu_{\delta}$ & $B_J$ & $R$ & $I$ & $J$ & $H$ & $K_s$ \\
SSSPM~J...   & (J2000) & & \multispan{2}{\hfil mas/yr \hfil} & \multispan{3}{\hfil (SSS) \hfil} & \multispan{3}{\hfil (2MASS) \hfil} \\
 \hline
 
1444$-$2019    & 14 44 20.67 $-$20 19 22.2 & 1998.351 & $-2919$\,$\pm$\,9 & $-$1950\,$\pm$\,5 & 21.159 & 18.569 & 14.954 & 12.546 & 12.142 & 11.933 \\ % 08_0024   
 \hline
 \end{tabular}

{\bf Notes:}
Coordinates are from 2MASS, which provides accurate positions at the most 
recent epoch. SSS magnitudes are mean values. The proper motion was determined 
from one 2MASS and six SSS positions. The addition of two DENIS measurements 
(mean $I$,$J$,$K$\,=\,14.945,12.485,11.891; epoch 2000.397) to the proper 
motion fit led to a less accurate solution: 
($-$2954,$-$1977)$\pm$(24,16)~mas/yr. 
\end{table*}
 
\normalsize

%%%%%%%%%%%%%%%%%%%%%%%%%%%%%%%%%%%%%%%%%%%%%%%%%%%%%%%%%%%%
%%%%%%%%%%%%%%%% PLOT OF 4 spectra %%%%%%%%%%%%%%%%
%%%%%%%%%%%%%%%%%%%%%%%%%%%%%%%%%%%%%%%%%%%%%%%%%%%%%%%%%%%%
%
\begin{figure}[htb]
\begin{center}
\includegraphics[width=6.1cm, angle=270]{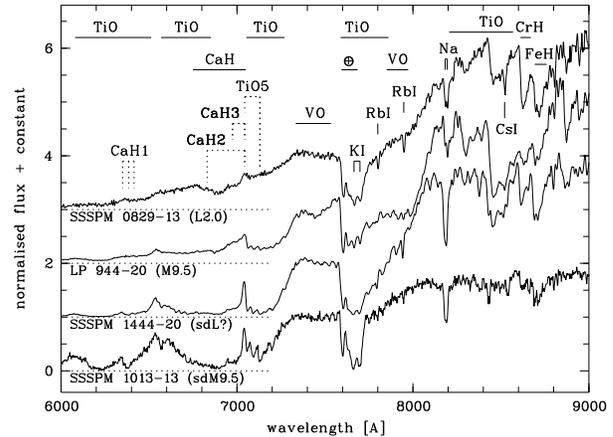}
\caption[4 spectra]{Optical spectra taken with EFOSC2 on the ESO\,3.6-m 
telescope for SSSPM~J1444$-$2019, LP\,944-20, and SSSPM~J0829$-$1309, along
with an NTT/EMMI spectrum of the latest known M-type subdwarf (sdM9.5) 
SSSPM~J1013$-$1356 (Scholz et al.\ 2004). The spectra are normalised at 
7500\AA{}. Key spectral features of M and L (sub)dwarfs are labeled. Also 
shown are the spectral regions defining the spectral indices TiO5, CaH1, CaH2,
and CaH3 used in the Gizis (1997) subdwarf classification scheme.
Uncorrected telluric absorption is indicated by the circled plus sign.
}
\label{4spec}
\end{center}
\end{figure} 

%__________________________________________________________________

\section{Spectroscopy}
\label{specobs}
Low-resolution optical spectroscopy was obtained on 16 March 2004 for 
SSSPM\,J1444$-$2019 and two comparison objects (spectral types M9.5 and L2) 
with EFOSC2 mounted on the ESO\,3.6-m telescope in service mode (ESO 
programme 072.C-0630). Using grism\#5, the wavelength coverage was 
5200--9350\AA{} with a FWHM resolution of $\sim$12.8\AA{}. 
The exposure times were 900s, 480s, and 600s for SSSPM\,J1444$-$2019,
LP\,944-20, and SSSPM\,J0829$-$1309, respectively. The spectral 
extraction, wavelength, and flux calibration were made using standard MIDAS 
routines. These spectra are shown in Fig.~\ref{4spec} along with an additional 
comparison spectrum (spectral type sdM9.5) obtained by Scholz et al.\
({\cite{scholz04}) observed with EMMI on the ESO NTT in May 2003. The 
resolution of the latter spectrum was $\sim$5\AA{} over the wavelength 
range 3850--9100\AA{}.

High-resolution (R\,$\sim$\,16,000 at 6000\AA{}) optical (5850--7400\AA{}) 
spectroscopy of SSSPM\,J1444$-$2019 was obtained using FORS2 on the ESO VLT,
along with a corresponding spectrum of LP\,944-20 (Tinney~\cite{tinney98a}) to 
serve as a comparison object (Fig.~\ref{2spec}). The observations 
were carried out in service mode on two different nights (29 August 2003 and 
17 December 2003) using grism GRIS\_1200R$+$93 with a slit width of 1~arcsec: 
the seeing was better than 0.8 arcsec during both observations. 
The total integration times were
1800s and 840s for SSSPM\,J1444$-$2019 and LP\,944-20, respectively. Data 
reduction was performed using standard IRAF tools and arc lamp wavelength 
calibration yielded an accuracy better than 0.2\AA{}. 

The FORS2 service observing scheme did not include the measurement of radial 
velocity (RV) standards during the same nights as our observations, but we have 
been able to use the comparison spectrum of LP\,944-20, which has a known 
RV of $+7.4\pm1.3$\,km/s (Tinney \& Reid~\cite{tinney98b}). 
By cross-correlating the two spectra shown in Fig.~\ref{2spec}, we were
able to measure a relative RV for SSSPM\,J1444$-$2019 of
$-149.2\pm3.8$\,km/s with respect to LP\,944-20. After correcting for the 
Earth's orbital motion during the different observing nights, 
a heliocentric RV estimate for 
SSSPM\,J1444$-$2019 of $-156.3\pm8.8$\,km/s was obtained, with the final error
dominated by the 0.2\AA{} accuracy of our wavelength calibration.

The spectrum of SSSPM\,J1444$-$2019 is remarkable in several aspects. Overall, 
it is much redder than that of the latest known M-type subdwarf 
SSSPM\,1013$-$1356 (sdM9.5; bottom of Fig.~\ref{4spec}) and comparable to that 
of the field brown dwarf LP\,944-20 (Tinney~\cite{tinney98a}), for which we 
adopted the spectral type M9.5 determined by Reid et al.\ (\cite{reid02}). 
The TiO absorption bands are extremely strong, as are the CaH bands, 
indicating a cool subdwarf nature. The complete absence of the VO bands, 
strong CrH and FeH, and in particular the presence of the atomic 
Rb\,{\small{I}} lines, are typical of L dwarfs, as can be seen in the
spectrum of the L2 dwarf SSSPM\,J0829$-$1309 shown in Fig.~\ref{4spec}
(Scholz \& Meusinger~\cite{scholz02}). 

The higher-resolution spectrum of SSSPM\,J1444$-$2019 shows no Li absorption 
at 6708\AA{}, while it is clearly visible in the known brown dwarf 
LP\,944-20 when observed with the same instrument (Fig.~\ref{2spec}). There
is apparent H${\alpha}$ line emission in SSSPM\,J1444$-$2019, despite the fact
that it is not usually observed in cool subdwarfs. However, it is not quite at 
the right wavelength, being shifted by about $-$2\AA{} with respect to the
expected location after taking into account the RV, and this
line may in fact be a blend of other, unrelated spectral features. 

We have applied the Gizis~(\cite{gizis97}) classification scheme for K and 
M subdwarfs to the SSSPM\,J1444$-$2019 spectra. The spectral indices TiO5, 
CaH1, CaH2, and CaH3 were measured to be 0.08, 0.42, 0.11, and 0.23, 
respectively from the low-resolution spectrum, and 0.079, 0.473, 0.120, 
and 0.232, respectively, from the high-resolution spectrum. The TiO5 and 
CaH2 indices are so small that they fall outside the plotted range in 
Fig.~1 of Gizis~(\cite{gizis97}), while the CaH1 and CaH3 values are very
close to the limits. The CaH2 and CaH3 indices for SSSPM\,J1444$-$2019 
are very similar to those measured for the sdM9.5 object SSSPM\,1013$-$1356 
(Scholz et al.\ \cite{scholz04}), while the TiO5 index has the smallest ever 
measured value. Conversely, the SSSPM\,J1444$-$2019 CaH1 index is larger than 
that of SSSPM\,1013$-$1356, but this index is not very useful for the 
latest-type objects according to Gizis~(\cite{gizis97}).

Applying the classification scheme of Gizis~(\cite{gizis97}), which is formally
valid only to sdM7, to the extreme spectral indices of SSSPM\,J1444$-$2019,
we compute a spectral type of sdM9. On the other hand, the L-type features 
and the very red continuum mentioned above would also allow us to classify
this object as an early-L subdwarf, similar to LSR~1610$-$0040 (L\'epine, 
Rich, \& Shara~\cite{lepine03}). However, the Gizis~(\cite{gizis97}) scheme 
yielded only sdM6.0 for this latter source, completely inconsistent with its 
much later spectral energy distribution.

%%%%%%%%%%%%%%%%%%%%%%%%%%%%%%%%%%%%%%%%%%%%%%%%%%%%%%%%%%%%
%%%%%%%%%%%%%%%% PLOT OF 2 high-res spectra %%%%%%%%%%%%%%%%
%%%%%%%%%%%%%%%%%%%%%%%%%%%%%%%%%%%%%%%%%%%%%%%%%%%%%%%%%%%%
%
%                                                One column figure
%----------------------------------------------------------- 
   \begin{figure}[htb]
   \centering
\includegraphics[width=6.1cm, angle=270]{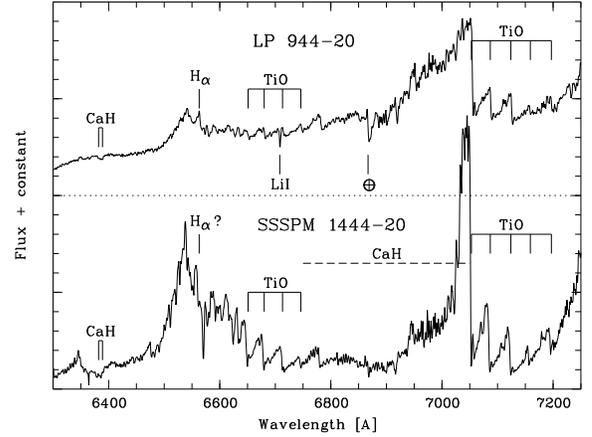}
\caption[2 high-res spectra]{FORS2 spectra of SSSPM\,J1444$-$2019 and the
brown dwarf LP\,944-20. The spectral region employed in the Gizis~(1997) 
subdwarf classification scheme is shown. The Li absorption line at 6708\AA{}
in LP\,944-20 and some of the sharp spectral features (mainly TiO) used in the
radial velocity determinations are marked. Telluric absorption is indicated 
by a circled plus sign.
}
\label{2spec}
   \end{figure}
%
%______________________________________________________________
 
\section{Discussion}
\label{concdisc}
Most of the known extreme HPM ($>$3.5 arcsec/yr) objects are very nearby 
($d$$<$10\,pc), the only exceptions being the DA7 white dwarf LHS\,56 at 
$\sim$16\,pc and the K-type halo subdwarfs LHS\,52/53, a wide binary
($\sim$8700\,AU separation) at $\sim$29\,pc. How far away is the newly
discovered HPM ultracool subdwarf SSSPM\,J1444$-$2019? 

With the discovery of an increasing number of ultracool subdwarfs, mostly 
via HPM surveys, it has become possible to explore an extension of the
subdwarf classification scheme into the late-M, L, and T dwarf regimes (see, 
e.g., Burgasser, Kirkpatrick, \& L\'epine \cite{burgasser04b}), but as yet, 
there is no formal scheme which covers SSSPM\,J1444$-$2019. Thus it is hard
to predict its intrinsic luminosity and hence distance; in addition, there 
are no trigonometric parallaxes available for similar objects, making direct
bootstrapping difficult.

For a first guess, we can compare the $I-J$ and $J-K_s$ colours of 
SSSPM\,J1444$-$2019 with those predicted by models of 8--10\,Gyr old, 
low-metallicity objects (Baraffe et al.\ \cite{baraffe97,baraffe98}). In 
Fig.~\ref{modeldata}, we see that the colours of SSSPM\,J1444$-$2019, the 
early-L subdwarf LSR\,1610$-$0040, and the late-L subdwarf 2MASS\,J0532$+$8246 
(Burgasser et al.\ \cite{burgasser03}) are all rather consistent with
moderately low metallicity models ([M/H]=$-0.5$). Concentrating on just
SSSPM\,J1444$-$2019 and LSR\,1610$-$0040, Fig.~\ref{modeldata} also shows that 
the models predict masses of $\sim$0.085 and 0.095\,M$_\odot$, respectively.
Then by comparing the model predictions for the absolute magnitudes for
sources with these masses with the measured apparent magnitudes for
SSSPM\,J1444$-$2019 and LSR\,1610$-$0040, we can calculate nominal distances
of $20\pm4$\,pc and $30\pm5$\,pc, respectively, conservatively including
0.4 magnitude uncertainties. Finally, we can then compute heliocentric space 
velocities following Johnson \& Soderblom (\cite{johnson87}), yielding 
$(U,V,W)=(-244,-256,-100)\pm(32,77,6)$\,km/s for SSSPM\,J1444$-$2019 and 
$(-60,-224,-79)\pm(15,41,9)$~km/s for LSR\,1610$-$0040. The space velocity 
of the former is clearly consistent with membership in the Galactic halo 
population, while that of the latter is typical of thick disk objects 
(Chiba \& Beers \cite{chiba00}).

% with corresponding absolute M$_I$, M$_J$, M$_H$, M$_K$ magnitudes of
% 13.54, 11.09, 10.51, 10.33 and 12.42, 10.55, 9.93, 9.67,

As discussed above, the formal classification of SSSPM J1444$-$2019 is sdM9,
although it appears likely that it is, in fact, an L-type subdwarf. The most 
convincing halo L subdwarf to date, 2MASS\,J0532$+$8246, has a characteristic 
late L-type optical spectrum, but with unusually strong TiO bands and a blue
near-infrared colour, remarkably similar in sum to SSSPM\,J1444$-$2019.
Burgasser (\cite{burgasser04a}) has also reported the discovery of another 
L subdwarf, 2MASS\,J1626$+$3925, which seems to extend the normal subdwarf 
sequence (with metallicities $-1.5<[\rm{M}/\rm{H}]<-1.0$) beyond 
SSSPM\,1013$-$1356 (sdM9.5) into the L regime (Fig.~\ref{modeldata}).

%%%%%%%%%%%%%%%%%%%%%%%%%%%%%%%%%%%%%%%%%%%%%%%%%%%%%%%%%%%%
%%%%%%%%%%%%%%%% PLOT OF models and data %%%%%%%%%%%%%%%%
%%%%%%%%%%%%%%%%%%%%%%%%%%%%%%%%%%%%%%%%%%%%%%%%%%%%%%%%%%%%
%
%                                                One column figure
%-----------------------------------------------------------
   \begin{figure}[htb]
   \centering
\includegraphics[width=6.3cm, angle=270]{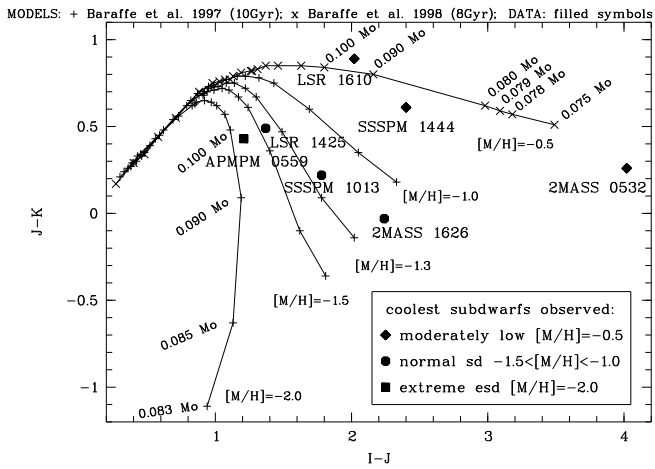}
\caption[models and data]{Colour-colour diagram for the coolest known subdwarfs
compared with evolutionary models (Baraffe et al.\ 1997, 1998). 
SSSPM\,J1444$-$2019, LSR\,1610$-$0040, and 2MASS\,J0532$+$8246 all appear
more consistent with moderately low-metallicity models than with their normal 
or extreme subdwarf counterparts.
}
\label{modeldata}
   \end{figure}

As one of the highest proper motion sources discovered in recent years and
probably one of the nearest ultracool sub\-dwarfs, SSSPM\,J1444$-$2019 should
be the subject of detailed follow-up observations and is a promising target 
for trigonometric parallax programmes. Further investigations of this kind 
are planned, including a comparison of our higher-resolution spectrum with 
model spectra in order to further quantify the metallicity and chemical
composition of SSSPM\,J1444$-$2019. 

%__________________________________________________________________

\begin{acknowledgements}
This research is based on data from the SuperCOSMOS Sky Surveys (SSS)
at the Wide-Field Astronomy Unit of the Institute for Astronomy, University
of Edinburgh. We have also used data products from the 2MASS, a joint project 
of the University of Massachusetts and the Infrared Processing and Analysis
Center/California Institute of Technology, funded by the NASA and NSF, and 
from the DENIS\@. We would like to thank Linda Schmidtobreick and Ivo Saviane,
the observers who took our ESO 3.6-m low-resolution spectroscopic observations 
in service mode, and the EFOSC2 support astronomer, Gaspare Lo Curto, for 
their efforts. We would also like to thank Mario van den Ancker for his 
advice during Phase~II preparations for our higher-resolution VLT FORS2
service observations and for his support in their scheduling. Finally, we 
thank Isabelle Baraffe and Gilles Chabrier for very helpful discussions. 

\end{acknowledgements}

%__________________________________________________________________

\end{document}